# Bulk-like Mott-Transition in ultrathin Cr-doped V$_2$O$_3$ films and the influence of its variability on scaled devices

*Johannes Mohr\*, Tyler Hennen, Daniel Bedau, Rainer Waser, Dirk J. Wouters*

J. Mohr, T. Hennen, R. Waser, D. J. Wouters
Institut für Werkstoffe der Elektrotechnik II, RWTH Aachen University, 52074 Aachen, Germany
E-mail: mohr@iwe.rwth-aachen.de

R. Waser
Peter Grünberg Institute, Forschungszentrum Jülich GmbH, 52428 Jülich, Germany

D. Bedau
Western Digital San Jose Research Center, 5601 Great Oaks Parkway, San Jose, CA 95119
E-mail: daniel.bedau@wdc.com

Keywords: Mott-transitions, doped V$_2$O$_3$, high-pressure measurements, percolation, scaling laws

The pressure driven Mott-transition in Chromium doped V2O3 films is investigated by direct electrical measurements on polycrystalline films with thicknesses down to 10 nm, and doping concentrations of 2%, 5% and 15%. A change in resistivity of nearly two orders of magnitude is found for 2% doping. A simulation model based on a scaling law description of the phase transition and percolative behavior in a resistor lattice is developed. This is used to show that despite significant deviations in the film structure from single crystals, the transition behavior is very similar. Finally, the influence of the variability between grains on the characteristics of scaled devices is investigated, and found to allow for scaling down to at least 50 nm device width.

## 1. Introduction

Recently, there has been significant interest in correlated electron materials for nanoelectronic applications.[1–7] Many of them exhibit phase transitions that drastically alter their electrical and often structural properties. Exploiting these transitions for devices, such as selector devices for resistive memories,[8–10] nanoscale oscillators for reservoir computing[11] or for neuromorphic computing,[12,13] could lead to major improvements in energy efficiency and integration density.

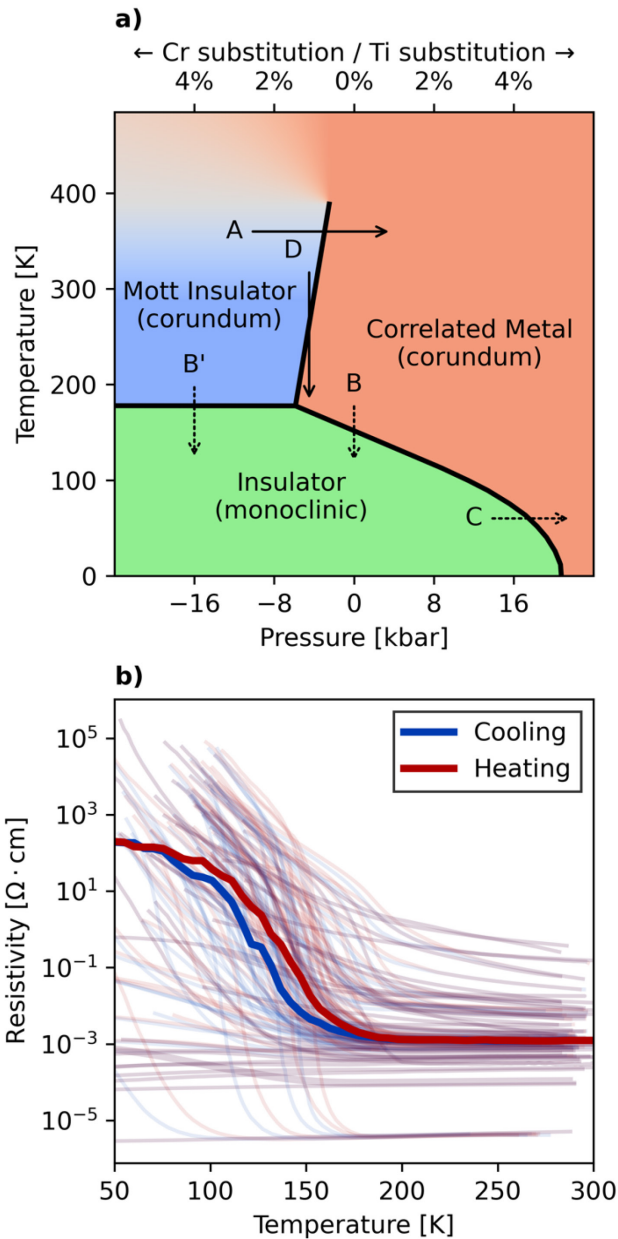

**Figure 1.** a) The phase diagram of Cr:$V_2O_3$. Chromium doping and pressure have opposite effects, with 1% doping equivalent to ~-4 kbar. b) Comparison of 85 literature measurements for the temperature driven phase transition in $V_2O_3$ (transition B), highlighting the wide range of resistivities and transition temperatures resulting from different stoichiometries and strains. The bold lines show the median value at each temperature for clarity.

However, for many materials, there has been significant debate regarding the underlying mechanism of the transition, and whether it is in fact related to correlated electron physics.

For example, in $NbO_2$ it was proposed that heating of a device triggers the material's phase transition, leading to a switching event. Others argued that temperatures required for this are too high to be plausible and proposed a thermal runaway effect as the origin.[14,15] Finally, it was shown that both mechanisms can be demonstrated in the same device.[16]

Here, we seek to elucidate the behavior of the Mott-transition in ultra-thin Cr doped $V_2O_3$ ($Cr:V_2O_3$) films. This material is of great interest for both theoretical and practical reasons. In bulk $V_2O_3$, doping with chromium triggers a phase transition that is seen as a prototypical Mott transition;[17] a purely electronic effect that does not involve any change in crystallographic phase. This leads to an increase in resistivity by many orders of magnitude. Application of pressure can reverse this transition and return the material to a metallic, conductive state.[17,18] This transition is of great practical interest for use in devices, because unlike the temperature driven transitions of materials such as $VO_2$ and $NbO_2$, there is no need to heat the device externally or through Joule heating to trigger a switching event. [16,19] Heating is not desirable from an energy efficiency standpoint, and requires the transition temperature to be within certain bounds. For example, the transition temperature in $VO_2$ is very close to room temperature,[20] devices therefore would not be usable in automotive or other challenging environments with elevated ambient temperatures. On the other hand, in $NbO_2$, such high temperatures are required that it is questionable whether it can be used in a device with reliable performance.[21]

In devices made from chromium doped $V_2O_3$, purely electronic switching has been reported, due to a proposed electronic avalanche mechanism,[22] which triggers a transition from the insulating to the metallic phase. On the other hand, the switching effect has also been claimed to be due to a self-heating effect,[23] as in $NbO_2$. Both non-volatile as well as volatile switching behavior has been found.[24]

Resolving this question is made difficult by the lack of any direct measurement of the materials resistivity across the pressure driven phase transition in thin films. It is known that the behavior of thin films can be significantly different from the bulk behavior, e.g. for undoped $V_2O_3$, strain due to the substrate can drastically alter its electrical behavior,[25,26] and the formation of metastable phases, not present in the bulk, has been reported.[27,28] Other potential influences could be the formation of interface layers, or the known difficulties in controlling the stoichiometry of $V_2O_3$ thin-films.[29] To illustrate this point, we have collected 85 measurements of the temperature driven phase transition in $V_2O_3$ for various thin films (**Figure 1b**)).[24,25,29–43] Clearly, the behavior is extremely variable, with e.g. the resistivity at room temperature covering a range larger than three orders of magnitude. It also significantly deviates from the bulk case, with the transition being altered in shape or height, or in some cases totally suppressed. This clearly shows that it cannot be assumed that the pressure driven transition in Cr:V$_2$O$_3$ thin-films behaves at all like the bulk.

It is therefore critical to characterize this behavior, especially because device applications typically require very thin layers, both for technological reasons and to achieve the high electric field strengths necessary to induce switching. A further complication is that films for use in devices cannot be grown epitaxially on a matched substrate, but will be integrated with other functional layers. This leads to polycrystalline growth and strain due to the different coefficients of thermal expansion of the film and its environment. Therefore, it might be expected that these devices suffer from significant variability if scaled to small dimensions. Here, we present results of electrical transport measurements on Cr:V$_2$O$_3$ thin-films with thicknesses down to 10 nm under hydrostatic pressures up to 15 kbar. We then develop a description of the bulk phase transition by a scaling law, and show via simulations that this also explains the observations in thin films, when variability is taken into account. Finally, the scaling limits for devices imposed by the polycrystalline growth are explored.

Some preliminary remarks regarding the rich phase diagram of Cr:V$_2$O$_3$ are necessary, to clarify how this work falls within the existing knowledge. Undoped V$_2$O$_3$ is, at room temperature, metallic with a corundum type structure. Upon cooling, it exhibits a transition to a monoclinic insulating phase, at approximately 150 K,[44] (Fig 1. a) – transition B). This transition has been well characterized also in thin-films,[42] and demonstrated in devices.[45] However, due to the required cryogenic conditions, these have limited practical applications. Doping V$_2$O$_3$ with ~2% chromium leads to another, distinct phase transition. This also results in an increase in resistivity by many orders of magnitude, but contrary to the low temperature transition, it is not accompanied by a change in crystal structure. Only a change in the lattice constants is observed, characterized by an increase in cell volume and a decrease in c/a-ratio. Applying a pressure of about 2 kbar returns the material to a conductive state and reverses the decrease in c/r-ratio[17,18,44](Fig 1. a) – transition A). This is the transition investigated here. Higher doping concentrations increase the required pressure to induce a transition, by about 4 kbar per percent Cr.[44] For very low doping concentrations, typically < 1.5 % it is also possible to cross this isostructural transition line by cooling (Fig 1. a) – transition D), reducing the temperature even further leads to a second, structural transition into the monoclinic phase (Fig 1. a) – transition B).

The doped material also exhibits a phase transition at low temperatures, changing from corundum to monoclinic structure, as in undoped V$_2$O$_3$ (Fig 1. a) – transition B').[18] Because this is a transition between two insulating phases, it is difficult to resolve in electrical measurements, and less interesting for applications. Finally, another pressure driven transition exists in undoped $V_2O_3$ at low temperatures. Because the transition temperature decreases with applied pressure, for a fixed temperature, an increase in pressure can also lead to an insulator-metal transition (Fig 1. a) – transition C). This has been investigated by Valmianski et al. [46]

## 2. Results and discussion
### 2.1. Thin-film characterization

Thin-films with three different chromium concentrations (2%, 5%, 15%) were characterized in this work. The former two are expected from literature results for bulk Cr:$V_2O_3$ to be close to the phase transition (see Figure 1 a)), the last one is investigated because it has been successfully employed to fabricate switching devices.[8,22,23] Therefore, it needs to be tested whether these are closer to the transition than expected. All samples used in the high-pressure experiments were fabricated by reactive RF magnetron sputtering from V/Cr alloy targets onto oxidized silicon wafers. To achieve the desired oxidation state, a very small concentration of $O_2$ was added to the Ar process gas. The substrates were heated to 600 °C to achieve a crystalline growth.

Characterization results for these films are shown in **Figure 2**. As expected, the films grow in a polycrystalline microstructure, as observed in the cross-sectional TEM shown in Figure 2 a), this is further confirmed by the GI-XRD spectra in d). These show most expected peaks for the corundum type structure of $V_2O_3$, however comparing to a reference for $V_2O_3$ powder, it is noticeable that some peaks are missing, such as the (1 1 0) peak at approximately 36°, or are weaker than expected, such as the (1 1 3) peak at 41°. This seems to indicate a preferred orientation of the crystallites, with the c-axis roughly perpendicular to the surface. A first important deviation from the bulk material is seen for the two thickest 2% films. These show several additional peaks, for example at 55° or 46°, which are not observed for higher doping concentrations. They might be explained by the formation of minor amounts of the metastable bixbyite phase of $V_2O_3$,[27] which has been reported to occur in sputtered pure $V_2O_3$ films.[28] Therefore, it is not surprising that these peaks are only found for very low doping concentrations.

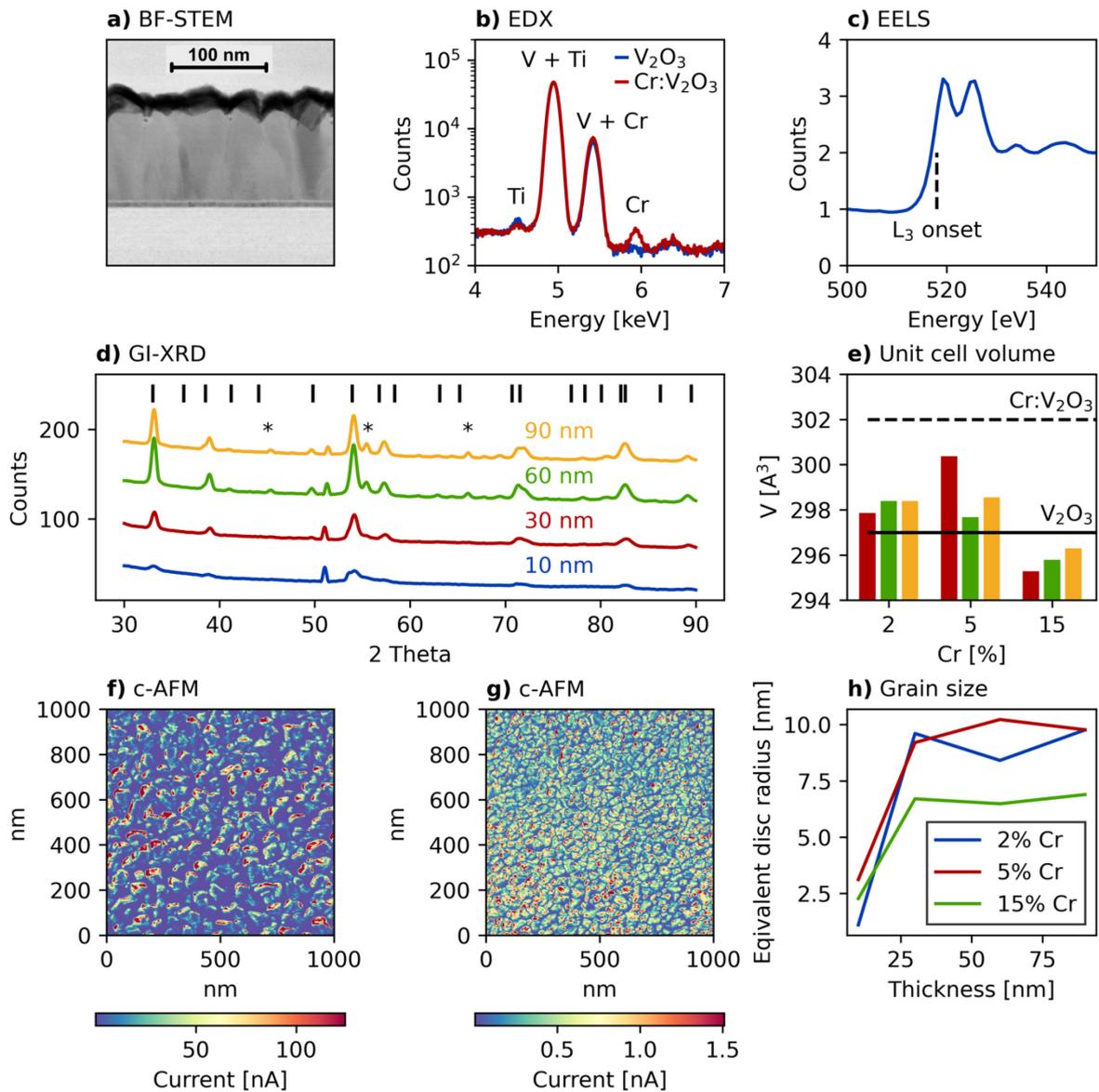

**Figure 2.** a) STEM image of a Cr: $V_2O_3$ film on TiN capped with Pt. b) EDX analysis of undoped and 2% Cr doped $V_2O_3$. c) EELS measurement of undoped $V_2O_3$. d) GI-XRD for different thicknesses of 2% doped films. Black lines indicate expected peaks for $V_2O_3$, stars mark peaks attributed to the bixbyite phase of $V_2O_3$. e) Unit cell volume extracted from GI-XRD measurements; colors correspond to the thicknesses in d) Lines indicate the literature cell volumes for single crystals. f) Conductive AFM mapping for undoped $V_2O_3$, g) for 15% Cr doping. h) Grain size extracted from non-contact AFM measurements.

The unit cell volume has been extracted from the GI-XRD measurements for different film thicknesses and chromium concentrations (Figure 1 e)). The addition of chromium leads to an expansion of the cell volume for bulk samples,[18] this is also observed here, however the effect is much smaller, with the volume still close the bulk value for $V_2O_3$. This might indicate a pre-existing strain in the material, potentially due to the growth conditions. The cell volume is found to be reduced for 15% doping (compared to the lower doping concentrations); this effect is known from the literature for nanoparticles.[47] Interestingly, there is also significant variation between different film thicknesses, with the 5% - 30 nm sample being an outlier closer to the bulk Cr:$V_2O_3$. This indicates that there is a very sensitive dependance of the internal strain state on the details of the growth conditions.

Regarding the film composition, Figure 2 b) compares a STEM-EDX for an undoped and a 2% Cr doped film. There is significant overlap between some peaks, however the isolated Cr-Kβ peak clearly indicates the incorporation of Cr into the film. This has been further confirmed by XPS, which shows that the peak intensity increases with doping concentration. The EELS result for an undoped sample (Figure 2 c)) is a further confirmation of the correct stoichiometry. The onset of the L3 transition accurately matches that reported by Kalavathi et al. [48]

Finally, another important distinction from the bulk material and epitaxial films is due to the grain structure. The c-AFM measurements in Figure 2 g, f) reveal that the material consists of rather fine grains with an inhomogeneous conductivity distribution. It is observed that, the interior of the grains is significantly more conductive, whereas the grain boundaries are insulating. It could be suspected that this is due to a segregation of Cr at the grain boundaries, however the effect also occurs in the undoped sample in g), clearly ruling this explanation out. It is more likely that this is also an effect due to an inhomogeneous strain distribution. These results suggest that the grain size might have a significant influence, therefore this is reported in h). Mostly, the grains are about 10 nm to 20 nm in diameter, which is in good agreement with the TEM image in a), except for the 10 nm thick films, which show much smaller grains. A high doping concentration also seems to reduce the size of the grains.

## 2.2. Characterization of the pressure driven phase transition

The high-pressure measurements were performed using a piston-cylinder type hydrostatic pressure cell manufactured by almax easylab. In such a cell, the sample to be measured is placed in a cavity filled with an (ideally) non-compressible fluid (**Figure 3 a**). A piston is then forced into the cavity, applying pressure to the medium, which in turn exerts an isotropic pressure on the sample. This type of cell can reach pressures above 35 kbar, [49] much less then for example diamond-anvil cells,[50] Due to the significantly larger volume however, it is much easier to accommodate the sample and the necessary electrical connections for the measurement. To prevent leakages of the pressure transmitting medium, it is enclosed within a Teflon cap and copper seal rings are used to fill gaps. Keeping this containment tight is the most challenging part of the procedure.

To fit into the cell, the samples were broken into smaller pieces, approximately 1 mm in length. Four short pieces of 100 μm wires were attached to the four corners of the pieces using conductive epoxy (Figure 3 a)). This epoxy offers a nominal conductivity of 0.0007 Ωcm, (doped $V_2O_3$: > 1 Ωcm), therefore the resistance of the connections should be insignificant compared to the Cr:$V_2O_3$ film.

The measurements were then performed by increasing the applied force by a small amount, measuring the cell pressure using a Manganin coil, and six values for the sample resistance. These are measured across both diagonals of the sample, as well as between all adjacent corners. The median of these values is reported. We have not measured according to the van-der-Pauw procedure, as the necessary conditions cannot be fulfilled due to the small sample size and required mechanical stability of the connections.[51] As the film resistance is much higher than that of the leads and contacts, this is not an issue. The pressure is then increased, and the measurement repeated, until a maximum pressure of around 15 kbar is reached. The maximum applied pressure varies between samples, as it is possible for the procedure to be interrupted by disconnected leads or chamber leaks.

Figure 3 b) summarizes the results. To show data for all chromium concentrations together, they are plotted over an "effective pressure". This describes the pressure needed to bring an undoped sample into an equivalent state. Because an expansion of the lattice is needed to do so, it typically is negative. This is calculated by assuming a pressure doping equivalence of 4 kbar per percent Cr ($k$):[44]

$$p_{\text{eff}} = p_{\text{meas}} - 4 \text{ kbar} \cdot k \tag{1}$$

The measured resistances are also scaled to account for the different film thicknesses ($t$):

$$R_{\text{norm}} = R \cdot t \tag{2}$$

Clearly, there is some pressure dependance for all doping concentrations. A significant jump in the resistance seems to be present, approximately at the border between the measurements for 5% and 2% doped samples. This is close to the effective pressure of -5 kbar where a phase transition would be expected from the bulk phase diagram in Figure 1 a). For 5% Cr, the resistance is relatively stable at low pressures, while at the highest pressures a significant drop in resistance is observed. For the 2% doped samples, a strong, but more gradual decrease in the resistance is seen, that occurs over the whole measured pressure range. Finally, for the 15% doped samples, a much weaker, gradual dependence on pressure is also measured.

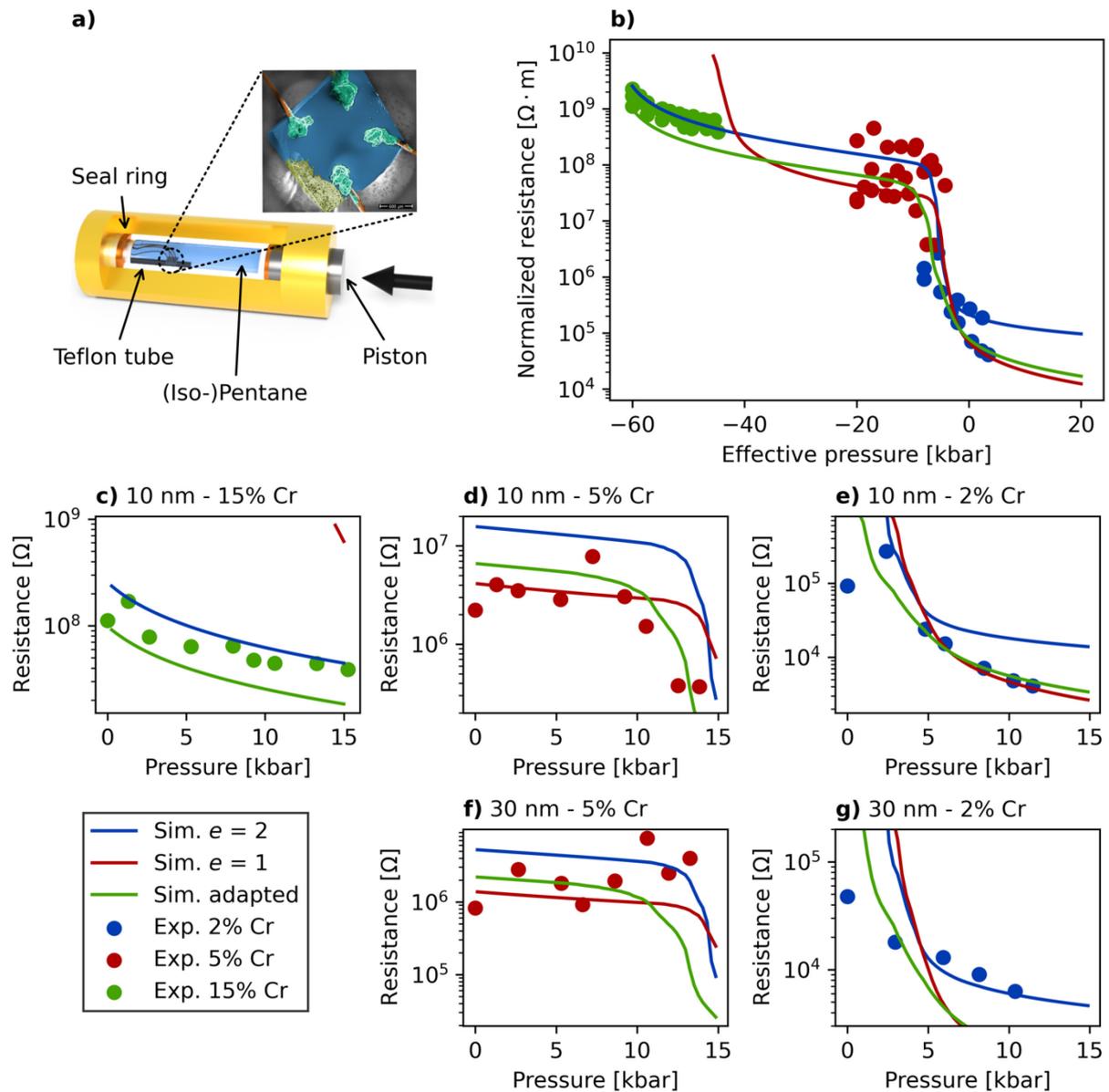

**Figure 3.** a) Sketch of the pressure cell used for the measurements, with SEM image of a sample with attached leads. b) Circles indicate measurement results for the resistance scaled with film thickness. Colors indicate the different doping concentrations. For 15% doping, ambient pressure corresponds to an effective pressure of ~-60 kbar, this is then increased during the measurement to about -45 kbar. 5% doping covers the range from -20 kbar to -5 kbar, while 2% doping ranges from -5 kbar to 10 kbar. Lines indicate simulation results for different models. c-g) Measured resistances over applied pressure for different samples.

To make the results clearer, in c-e) the measurements for 10 nm thin films are shown over the actual applied pressure and the real resistance is given. For 15% doping, the highest resistance is observed, this decreases about half an order of magnitude over the observed pressure range of ~15 kbar. For 5% Cr, the resistance is smaller, and first shows a constant plateau, before showing an abrupt drop in resistance above 10 kbar. At 2%, the resistance is even further reduced, and a gradual decrease in resistance is seen, covering nearly two orders of magnitude over the observed range. This sequence of concentrations corresponds to gradually approaching the transition line (see Figure 1 a))

In the case of 30 nm films (Figure 2 f, g), for 2% doping, a somewhat less pronounced reduction is seen (about an order of magnitude). Interestingly for 5% doping there is no drop in the resistivity, it even appears to be slightly increasing.

Finally, it must be considered that irreversible changes in the material might occur, which could be mistaken for a phase-transition. To rule this out, it is necessary to also measure at decreasing pressures, which is challenging because it is challenging to control a pressure decrease in our clamp cell, and for many experiments the seals failed once the highest pressure was reached. Nevertheless, we have been able to collect some data for these conditions (**Figure S1**). A clear tendency for the resistance to return to its initial values is observed. The data for loading and unloading do not match perfectly, which might be due to the hysteresis observed in the phase transition of bulk samples, however an influence of inaccurate pressure readings cannot be completely ruled out.

## 2.3. Modeling of the phase transition in thin-films

The presented results clearly indicate that the films show a strong pressure dependence of resistivity, but this cannot immediately be accepted as proof of the Mott transition, as there are other phenomena that can lead to a change in resistivity with pressure. In fact, many materials show this to a certain degree,[52] and the strain dependence of the conductivity of silicon[53] finds many technical applications in MEMS,[54,55] as well as to improve the properties of transistors.[56–58] Bridgman presented data for various other oxides under hydrostatic pressure,[52] which we have compared to our data in **Figure S2**. This clearly shows that the pressure dependence of the 2% and 5% doped films is much higher, as expected for a phase transition, whereas for 15% it is on the same order of magnitude as for other materials. To enable a more quantitative comparison of our measurements to the bulk behavior, we have developed a simulation model. Our approach considers two superimposed phase transitions, the pressure driven Mott-transition in the individual grains, and the percolative transition of the system of grains. First, we seek an adequate description of the behavior in single crystals. It is then assumed that the individual crystallites in our thin-films show the same fundamental behavior, but follow a distribution of transition pressures due to different internal strain states. As it has been reported before that the phase transition in Cr:V$_2$O$_3$ can be described by a universal scaling law,[59] we consider a model of the following form:

$$\sigma(P) = \begin{cases} -a \cdot (P_C - P)^{\frac{1}{e}} + \sigma_C, P < P_C \\ b \cdot (P - P_C)^{\frac{1}{e}} + \sigma_C, P > P_C \end{cases} \quad (3)$$

Where $P_C$ is the critical pressure of the transition. Separate pre-factors were used for each side of the transition, and the critical exponent $e$ was the same.

This model was then adapted to the measurements by McWhan on single crystals doped with ~4% Cr.[18] These contain both a branch for increasing and decreasing pressure. The former was selected to match the experimental conditions used here. These data show a different behavior depending on the crystal orientation, and both are shown together with the adapted models in **Figure 4 b)**. Clearly, the model can describe the data extremely well, but different critical exponents must be used for the different crystal orientations.

Both values are plausible from the literature on other metal-insulator transitions. For the doping driven transition, values close to $e = 1$ are found in many studies of doped Si and Ge,[60] as well as in amorphous $Nb_xSi_{1-x}$.[61] On the other hand, $e = 2$ is measured for example in P doped Si.[62] It must be noted however, that the influence of doping in these materials is to increase conductivity in contrast to the case of $Cr:V_2O_3$. Finally, these values are also among those predicted by the scaling theory of localization.[63]

As a second step, the model considers the polycrystalline nature of the films, and the clearly inhomogeneous electrical properties observed. It is reasonable to assume that this will lead to a transition that proceeds as a percolative process, with grains undergoing the phase transition at slightly different pressures. This is also justified by reports that the phase transitions in both $Cr:V_2O_3$[64] and undoped $V_2O_3$[65] are percolative processes even in single crystals and epitaxial films respectively.

To take this into account, the film is modeled as a 2D network of hexagonal cells (Figure 4 a)), somewhat similar to the approach taken by Stoliar et al.,[22] the resistivity of each one being described by Equation 1. These were chosen due to the hexagonal symmetry of the unit cell of $V_2O_3$ and the texture of the films. The exponent and pre-factors are the same for all cells, however the critical pressure is randomly selected from a normal distribution with a specified mean ($p_{th}$) and standard deviation ($p_{var}$). Each cell contains six resistors of equal value that connect its center point to the neighboring cells. The resistors then form a triangular lattice dual to the hexagonal one (lattice type $3^6$), with a percolation threshold known to be 0.5.[66] Because the resistor values are finite and continuous, no discontinuity in the percolation phase transition is expected, but qualitatively similar behavior. Finally, a scaling factor is applied to all resistors in the network. This is necessary because the model is 2D, requiring the data to be scaled to account for film thickness and because the simulated lattice cannot represent the physical sample size due to computational cost (as there are more than a billion grains in a 1 mm × 1 mm sample).

To produce a resistance over pressure curve, the current through the network is then solved for each pressure with the resistors set to the appropriate values. Figure 4 c-e) shows the resistance states in a 20 × 20 network of cells across the phase transition. At low pressures (c)), most cells are highly resistive, with only a few conductive ones interspersed. At the transition pressure (d)), approximately half of the cells are conductive, consistent with the expected threshold for a pure percolation transition. Finally at high pressures, only a few highly resistive cells are left in a conductive environment.

Figure 4 f-h) shows the potential distribution in the network when a voltage is applied to the top interface and the bottom interface is grounded. The calculated potential is transversely homogeneous before and after the transition, whereas near the transition pressure a pronounced structure is observed.

With these simulations, it can be shown that the observed behavior in thin-films is consistent with a bulk like Mott-transition occurring in the individual crystallites. We minimized the number of fitting parameters to avoid ambiguity in the results; the resistance scaling factor was adapted to our data to bring the absolute resistance values of experiment and simulation into agreement, and $p_{var}$ was also varied. $p_{th}$ is calculated assuming that the pressure doping equivalence of 4 kbar per percent chromium holds. Then the critical pressure of the adapted models can be converted (knowing their doping level of 4%) to the expected effective $p_{th}$ for an undoped sample (approximately -5 kbar). A $p_{var}$ of 1.5 kbar appears to describe a variability consistent with the experimental data.

The simulated behavior is compared to the measurement data in Figure 3b). For both models, good agreement of both the transition pressure and the magnitude of the resistance change is achieved between measurements and simulations. Considering the high-pressure behavior above the transition, it appears the experimental data shows two different slopes. These correspond to 10 nm and 30 nm thick, 2% doped samples. These are shown individually in Figure 3 e,g). It appears like the model with $e = 1$ is a good description of the 10 nm film, while $e = 2$ adequately describes 30 nm. Given that the two models are for measurements along different crystallographic orientations, it seems plausible that the thin-films grow with different textures depending on their thickness, and one or the other model is more appropriate.

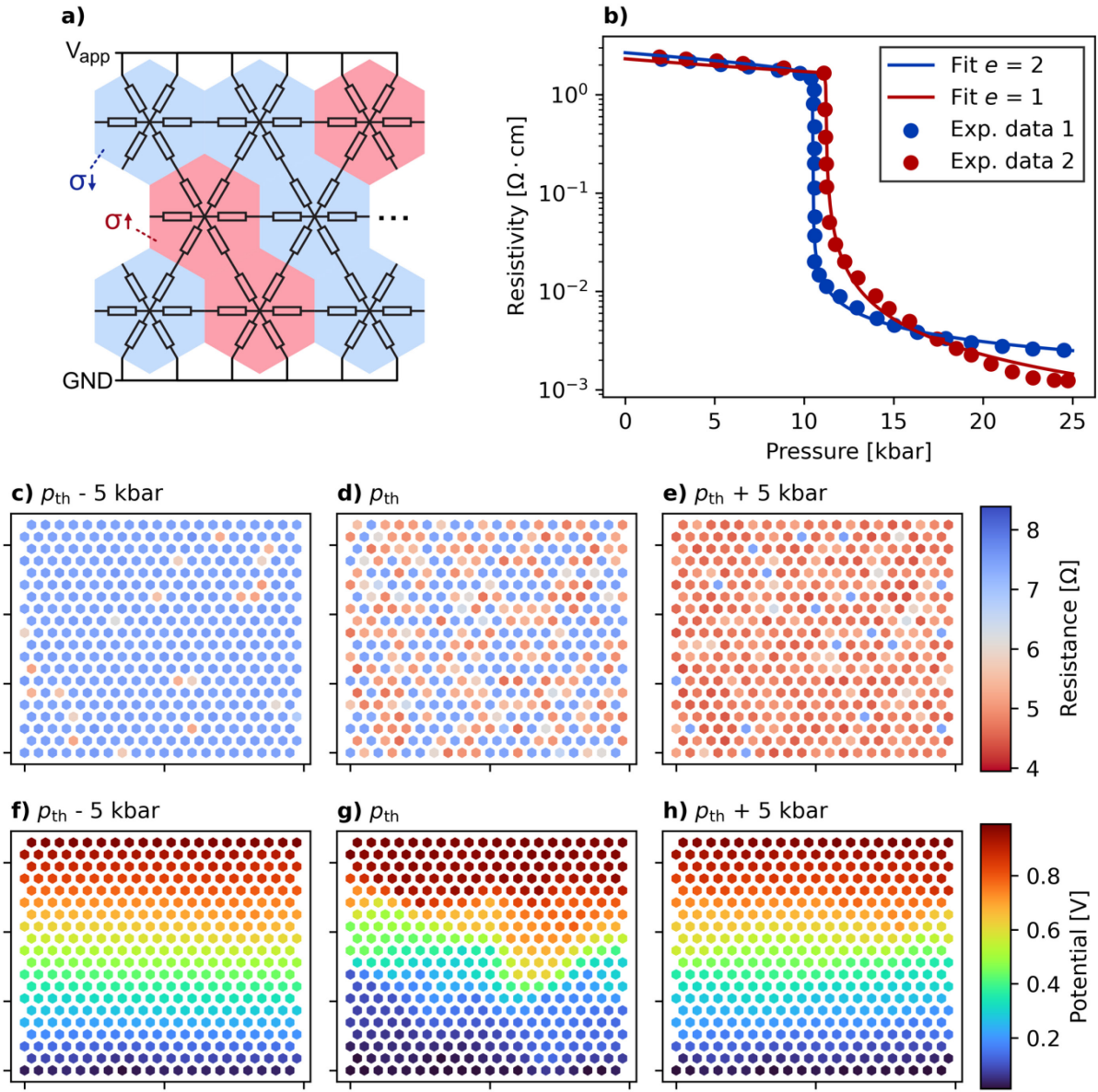

**Figure 4.** a) Sketch of the resistor lattice used for the simulations. b) Measurements of the Mott transition in single crystalline Cr:V$_2$O$_3$ (circles) reported in the literature,[18] with fit scaling law models (**Equation 3**). c-e) Example of the cell resistances across the phase transition for a 20 × 20 lattice. The model with $e = 1$ was used, with a standard deviation of 3 kbar. f-h) The potential distributions in the lattice corresponding to the resistance states.

Considering the 5% doped samples, corresponding to effective pressures right at the transition, for the 10 nm film (Figure 3 d)) it seems that the models also predict the onset of the phase transition well, however the actual transition occurs at slightly lower pressures than expected. This is also found for a 90 nm sample (only shown in Figure 3b)). On the other hand, for the 30 nm case (Figure 3 e)), no transition is observed in the experimental data. This is most likely a shift in the transition pressure to higher values, so it is no longer within the measurement range. These variations can be understood as an effect of pre-existing strain in the films, as they are consistent with the cell volumes found by GI-XRD. These are smaller than expected from bulk samples, corresponding to a unit cell that is already slightly compressed, and therefore less external pressure is needed to induce a transition. This explains the reduced pressures needed to induce the transition in 10 nm and 90 nm case. Interestingly, by far the largest unit cell is found for the 30 nm film, so it reasonable to assume that the strain state here is such that the transition is close to its bulk value, or shifted to slightly higher values. Then, it is not surprising that the transition occurs at higher pressures, outside the range of this measurement.

Finally, the pre-transition behavior must be considered. This is observed for very low effective pressures, corresponding to the 15% doped samples. Clearly, the model with $e = 1$ is not a satisfactory description in this range, whereas with $e = 2$ a remarkably good agreement in both the resistance values as well the slope over pressure is found.

While the reasons for specific values cannot be fully elucidated here, it should be considered that the exponents on both sides of the transition do not necessarily need to be the same. A sharp metal-insulator transition can only be observed at T = 0 K,[67] because thermal activation will otherwise lead to some conductivity in the insulating phase. In the ideal transition, the conductivity behaves like an order parameter and vanishes on one side of the transition.[67] For Cr:V$_2$O$_3$ nanoscale devices, an exponential temperature dependence was observed.[8] It is then plausible that the real behavior in the insulating phase is governed by different physics, requiring a different mathematical description. Unfortunately, low temperature measurements to clarify this cannot be performed in the case of Cr:V$_2$O$_3$ because the crystal phase of interest only exists at relatively high temperatures (see Figure 1 a)).

Finally, an adapted model was created to describe the 10 nm films as accurately as possible, to enable predictions regarding device behavior. It takes into account the shift in transition pressure, as well as the two different exponents:

$$\sigma(P) = \begin{cases} -a \cdot (P_C - P)^{\frac{1}{2}} + \sigma_C, P < P_C \\ b \cdot (P - P_C) + \sigma_C, P > P_C \end{cases} \quad (4)$$

The transition pressure was shifted by -2 kbar compared to the value predicted by the pressure doping equivalence, and $p_{var}$ = 3 kbar was assumed. Figure 3 c-d shows that this model is in excellent agreement with the observations.

## 2.4. Influence of the variability on scaled devices

If it is assumed that there is a relevant variability in the critical pressures between different grains, it must be investigated what influence this has on the observed macroscopic resistivity. Especially in the case of scaled devices, where only a relatively small number of grains are involved in the switching, this might lead to significant variability between them.

To investigate this, we repeated the resistor lattice simulations a number of times, with the transition pressure of each cell randomized again before every run. **Figure 5** shows the results for different numbers of cells, both in square configurations (Figure 5 a-c)), as well as for wider than long devices (Figure 5 d-e)), and vice versa (Figure 5 g-i)). The final adapted model was used, as well as a version with an increased $p_{var}$ = 5 kbar, to show more clearly the trends in device behavior.

The results for the square configurations show two main trends for increasing lattice size. First, the spread between the different curves is reduced, and second the observed transition becomes more gradual. This is expected, as for the 5 × 5 lattices, any single cell will have a large influence on the total resistance, if it experiences the phase transition, a significant jump will be observed. Because the number of cells is small, the mean transition pressure will also be different between simulation runs, resulting in a horizontal shift of the curves. On the other hand, for the 20 × 20 case, the average behavior of a large number of cells is measured, smoothing out the transition, and bringing the mean close to the specified critical pressure.

To estimate the influence of this variability on the device scaling limits, we focus on the most relevant 10 nm films. By AFM, a grain diameter of approximately 5 nm was found for these, largely independent of doping concentration (Figure 2 f)). Then the 10 × 10 lattice would correspond to a device size of 50 nm × 50 nm. For the model with 1.5 kbar standard deviation (which describes the experimental films), only minor differences between devices are observed, indicating that they could reliably be used in applications. On the other hand, for the 5 × 5 lattice, some transition curves are outliers, which would mean some devices show significantly different behavior.

Interestingly, for the asymmetric structures, increasing the lattice size does not always reduce the variation. For the wide devices, an increase in device width appears to also increase the variability at lower pressures. This is because they consist of only a few layers of cells vertically, if only a few cells close to each other undergo the phase transition, a conductive path through the device will be formed. If the device width is increased, so does the likelihood of a cluster of cells with very low transition pressures to be part of the layer. Because of this, the majority of the variability is observed before the average transition pressure.

On the other hand, for the case of long and thin devices, the high-pressure behavior is more variable. Here, a small number of cells transitioning very early does not have a significant influence on the total resistance, because they are followed by many layers of high resistance cells. However, if the majority of cells have completed the transition, then only a few cells with an extremely high critical pressure that remain highly resistive can interrupt or constrict the current flow.

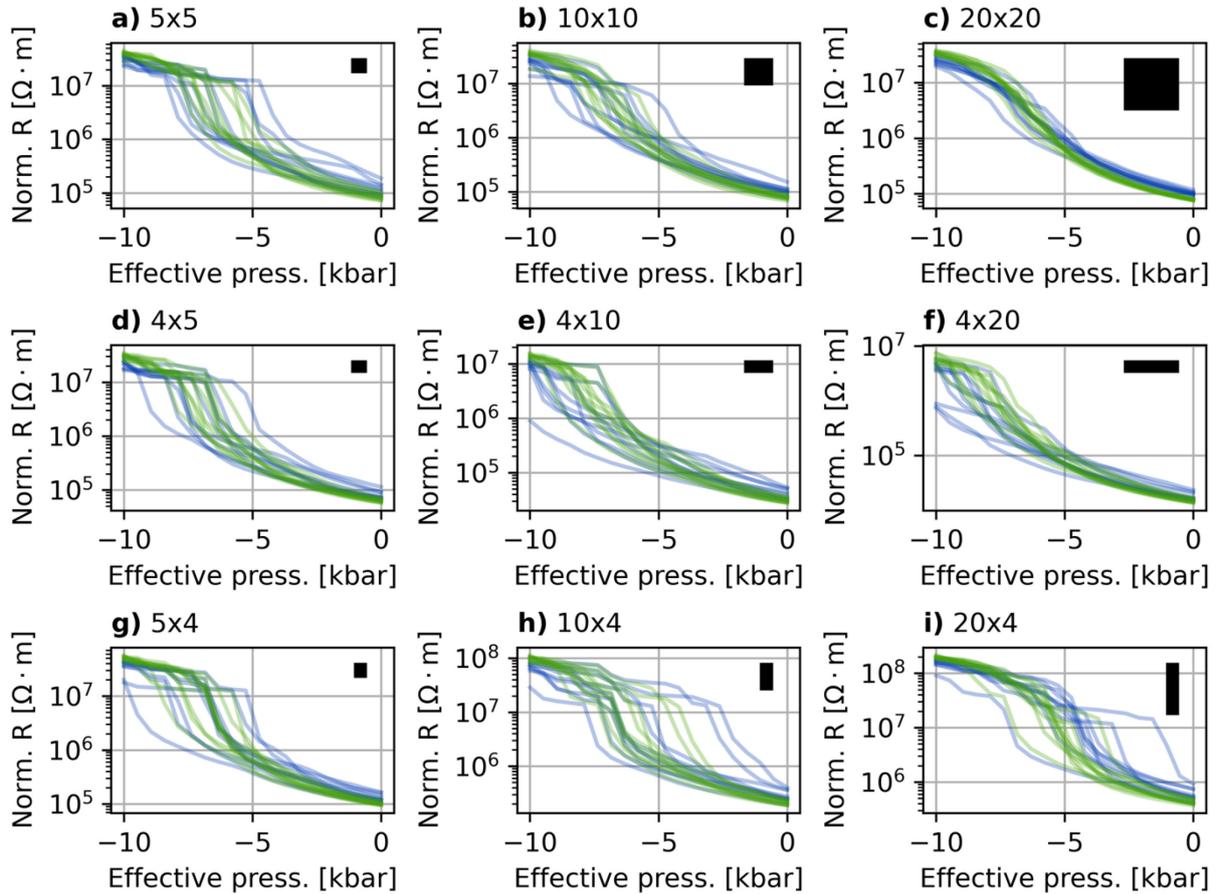

**Figure 5.** Simulation results for different lattice sizes. The model adapted for the 10 nm films was used (green), to demonstrate the differences more clearly, an increased standard deviation of 5 kbar is also shown (blue). The geometry of the lattice is indicated by the black box. In a-c) square lattices are shown, in d-e) wide and short structures and in g-i) long and thin ones.

## 3. Conclusions

The results presented above are strong evidence that a pressure driven Mott-transition occurs in films down to 10 nm in thickness. This is very promising for device applications, especially because the transition appears to be remarkably robust. The films depart from an ideal single crystal sample significantly, especially due to their polycrystalline grain structure with insulating grain boundaries, as well as being subject to strains from the growth conditions. Furthermore, it can be expected that at least the surface exposed to ambient atmosphere is not perfectly stoichiometric, and finally, for the measurements the samples are submerged in solvents. Nevertheless, the magnitude of the resistance change during the transition appears to be just as large as in the bulk, and its onset is only slightly shifted. The former result is most important, because a suppressed transition hampers device applications, whereas the position can be largely adjusted to the requirements by changing the doping concentration.

In addition, for practical applications, it is important to take into consideration the complex behavior of polycrystalline films. On the one hand side, there is a phase transition in the material itself, on the other hand the polycrystalline media also undergo a percolation phase transition as the individual grains switch. Frequently, individual grains will have a large stress distribution, leading to a $2^{nd}$ order broadening effect compared to single crystals.

As we have demonstrated, the variability strongly depends on the aspect ratio, an important design factor for nanostructures. It is however sufficiently low that devices could be scaled down to at least 50 nm × 50 nm without suffering from excessive device-to-device variation.

## 4. Methods

*Sample fabrication*: The samples for the high pressure and GI-XRD experiments were fabricated on 1" × 1" silicon wafers covered with a ~430 nm thick oxide layer. Those for the conductive AFM experiments were deposited on a 30 nm thick Pt layer. The wafers were cleaned in acetone and isopropanol in an ultrasonic bath for 10 minutes each. Then they were thoroughly rinsed in deionized water and dried with nitrogen. The samples were then transferred into the deposition system.

The $Cr:V_2O_3$ films were deposited by reactive magnetron sputtering using 1" V/Cr alloy targets with 2%, 5% and 15% nominal concentrations. The sputter sources were at an angle of 45° to the substrate holder, which rotates to facilitate a homogeneous deposition. An atmosphere consisting of Ar and $O_2$ was used to oxidize the films, the oxygen content in the mixture was 600 ppm. The process pressure was set to 0.010 mbar, the deposition power was 50 W. The substrates were heated by halogen bulbs on the backside to a deposition temperature of 600 °C before the deposition. After completion, they were allowed to cool to ~100 °C in the chamber before being removed from the system. The film thickness was controlled via the deposition time, to facilitate this, the deposition rate for each target was calibrated by x-ray reflectivity measurements.

As it was found that the properties of the films depend very sensitively on the state of the chamber, exemplified by the first deposited film per day being significantly different from the rest, it was preconditioned by performing a dummy run before any actual depositions. These were then done in immediate succession, so that the chamber temperature and residual gas concentrations were as reproducible as possible.

*High pressure measurements*: For the high-pressure measurements, the substrates were broken into smaller pieces, roughly 1mm to 1.5 mm in size. Short pieces of 100 μm thick copper wire were tinned on both ends. The epoxy (MG Chemicals 8330S) was mixed according to the manufacturer's specification and one end of the wires were dipped into it. These were then carefully attached to the four corners of the samples. The epoxy was cured on a hotplate at 80 °C to 90 °C for approximately one hour.

The feedthroughs for the pressure cell are provided with a sample platform and four wires, as well as a Manganin coil as a manometer. The samples were attached to the platform with a small droplet of glue, then the wires on the sample were soldered to those coming from the feedthrough.

The Teflon cap was filled with the pressure transmitting medium (pentane or an isopentane/pentane mixture) and the feedthrough was inserted into it. This assembly was then placed in the pressure cell and this was closed according to the manufacturer's procedure.

*Grazing incidence X-ray diffraction (GI-XRD)*: All measurements were done using a PANalytical X'pert Pro MPD diffractometer with Cu K α1 radiation at an incident angle of 0.4°. An alignment procedure for the sample height and tilt angle was performed. Due to the relatively weak signal from the films, a measurement time of 12 hours was used. The data shown in the figures were noise filtered using a Savitzky-Golay filter. Peaks were identified by comparing the pattern to the reference data from PDF 00-034-0187.[68] There is one peak in the data, at ~52° that is an artifact due to the sample holder.

*Atomic force microscopy (AFM)*: The measurements were done in a Park NX10 system. The structure of the films used for the pressure experiments and the grain size distribution was analyzed using OMCL-AC160TS cantilevers in non-contact mode. The conductive AFM images were recorded with AS-2.8-SS conductive diamond cantilevers in contact mode, with a bias voltage of 0.1 V applied to the sample. The samples were stored in vacuum to reduce moisture and contaminations of the sample surface as much as possible.

*Simulations*: Simulations were done using the Cadence Spectre simulator. A netlist was prepared corresponding to the hexagonal lattice of cells. All resistors in one cell had the same conductivity, which was calculated using the scaling law depending on the pressure, and the specific transition pressure of that cell. These were randomly selected from a normal distribution at the beginning of each run, and kept the same for all pressure steps. To evaluate the resistance, the topmost row of resistors was connected to a 1 V voltage source, the bottom row was grounded. Resistors on the left and right boarders of the lattice were left unconnected.

**Acknowledgements**

The authors acknowledge funding by the DFG (German Science Foundation) within the collaborative research center SFB 917.

The Mott-transition in thin Chromium-doped $V_2O_3$ films is characterized by electrical measurements under high pressure. Using simulations, it is confirmed to be very similar to the single crystal behavior down to 10 nm film thickness, despite an imperfect film quality. Finally, the implications of variability between grains for scaled devices are demonstrated.

J. Mohr*, T. Hennen, D. Bedau, R. Waser, D. J. Wouters

**Bulk-like Mott-Transition in ultrathin Cr-doped V2O3 films and the influence of its variability on scaled devices**

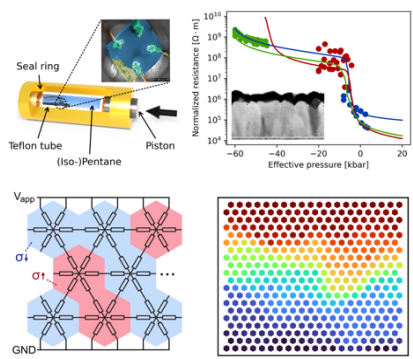

Supporting Information

**Bulk-like Mott-Transition in ultrathin Cr-doped V$_2$O$_3$ films and the influence of its variability on scaled devices**

*Johannes Mohr\*, Tyler Hennen, Daniel Bedau, Rainer Waser, Dirk J. Wouters*

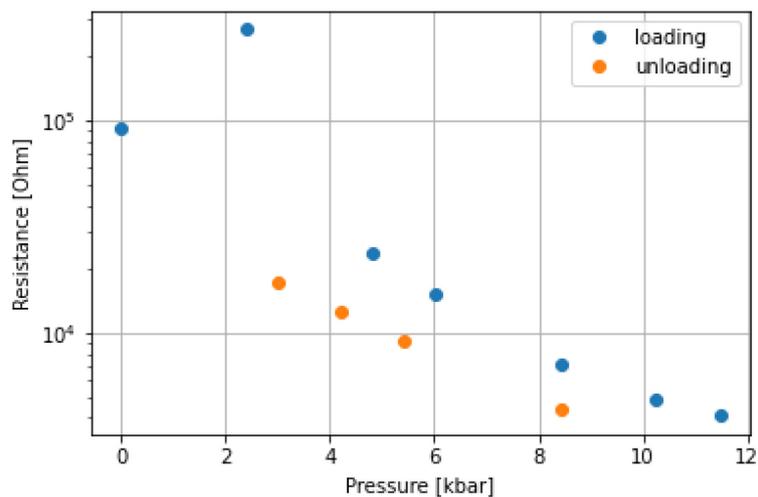

**Figure S1.** Comparison of resistance measurements on a 10 nm, 2% Cr film for increasing (loading) and decreasing (unloading) pressure. Some hysteresis is observed, but the resistance change appears to be reversible in principle.

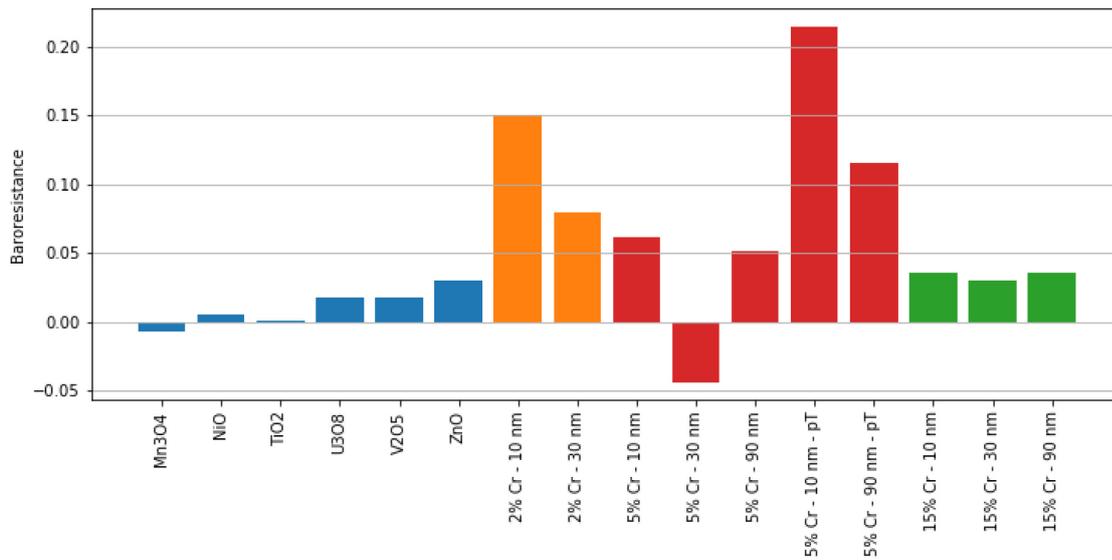

**Figure S2.** The relative resistance change in different films is compared to literature measurements on other oxides[52] (blue bars). For the measurements on 2% and 5% doped $V_2O_3$ the effect is calculated both over the whole pressure range, and selectively to the right of the transition (pT).